\documentclass[a4paper,12pt]{article}
\usepackage[dvips]{graphics}
\usepackage{a4wide}
\begin{document}


\title{Some remarks  on  the   Lieb-Schultz-Mattis theorem   and  its
	extension to higher dimensions}

\author{G. Misguich$^*$, C. Lhuillier\thanks{Laboratoire de Physique
	Th{\'e}orique des Liquides UMR 7600 of CNRS. Universit\'e P.
	et M.  Curie,  case 121, 4 Place  Jussieu, 75252 Paris Cedex.}
	}
\maketitle 
\bibliographystyle{prsty}


\abstract{
The extension of the Lieb-Schultz-Mattis  theorem to dimensions larger
than  one  is  discussed.    It is   explained  why   the  variational
wave-function built by the previous authors is of no help to prove the
theorem in dimension larger  than one. The  short range R.V.B. picture
of  Sutherland, Rokhsar  and Kivelson, Read   and Chakraborty gives  a
strong support  to the assertion that the  theorem is  indeed valid in
any dimension.  Some illustrations of  the general ideas are displayed
on exact spectra.
PACS numbers: 71.10.Fd; 75.10.Jm; 75.40.-s; 75.50.Ee; 75.60.Ej; 75.70.Ak
}

\maketitle
\begin{section}{Introduction}

In 1961    Lieb,   Schultz and   Mattis proved    that    a spin   1/2
antiferromagnetic  periodic chain  of length   $L$ has   a low  energy
excitation of   order  ${\cal O} (1/L)$~\cite{lsm61}.     This theorem
(called in the following LSMA)  was then extended  by Lieb and Affleck
to {\em odd-integer} spin but fails for  integer ones~\cite{al86}.  It
states that SU(2) invariant Hamiltonians with odd-integer spins in the
unit cell, either have gapless excitations or degenerate ground-states
in the thermodynamic  limit.  The authors  suggested that it might  be
extended  to   higher space dimensions, but  up   to now,  no complete
argument has been worked out~\cite{a88}.

In this paper we revisit  the method used by  LSMA (construction of  a
variational  excited state) and the   physical meaning of the  unitary
operator involved in this construction.   This sheds some light on the
reason why  the  LSMA excited state   is  generally not a   low energy
excitation in  space dimension larger than one  (as for example in the
case of N\'eel order on the triangular lattice), and how qualitatively
one might try to build a truly low energy excited state. We then study
an alternative wave-function very much in the spirit of the resonating
valence-bond (RVB)  states    of Sutherland~\cite{s88},   Rokhsar  and
Kivelson~\cite{rk88},  and  Read and Chakraborty~\cite{rc89}.  In this
last  framework, we show that the  above-mentioned statement is indeed
true in any dimension, and discuss the  quantum numbers of these quasi
degenerate ground-states.

\end{section}
\begin{section}{The LSMA Theorem}

 To analyze   LSMA theorem in   dimension larger than  one a dichotomy
 could  be done  between situations  {\em  with} $T=0$ {\em long-range
 order and  symmetry breaking   order  parameters} on one   hand,  and
 systems {\em without long-range order} on the other.

\begin{itemize}

\item In the first case the theorem is trivially true: a
symmetry-breaking  situation necessarily  implies in the thermodynamic
limit   the   mixing of  states  belonging   to  different irreducible
representations of  the Hamiltonian and/or  lattice symmetry group and
thus a degeneracy of the ground-state.

 {\em i)}  In the case  of N{\'e}el  long-range  order, there are both
 degenerate states (which   form the true  thermodynamic ground-state)
 and    gapless   excitations.     The    gapless  excitations,    the
 antiferromagnetic magnons, are the  Goldstone mode associated  to the
 broken continuous SU(2) symmetry.  Whereas it is  well known that the
 softest magnons scale as ${\cal O} (1/L)$, it  is often asserted that
 these quasi-particle first  excitations are the first excited  levels
 of the multi-particle spectra.  This is indeed false: the "T=0 N\'eel
 ground-state"    (or  "vacuum of  excitations")  is   itself a linear
 superposition  of $\sim   N^{\alpha}$   eigen-states of the   $SU(2)$
 invariant hamiltonian which collapse to the ground-state as ${\cal O}
 (1/L^2)$  ($\alpha$ is the   number  of sub-lattices of the  N{\'e}el
 state, $L$ is the  linear size of  the  sample and $N$ the  number of
 sites of the sample)~\cite{f89,nz89,bllp94}.

 {\em ii)} In the case of a discrete broken symmetry, as for example the
 space symmetry breaking  associated to long-range dimer or plaquette
 order,    or in  the   case of   $T$-symmetry breaking  associated to
 long-range  order in chirality,   the vacuum of  excitations  is also
 degenerate   in the    thermodynamic  limit.   The   number of  quasi
 degenerate states is related to the dimension  of the broken symmetry
 group, the nature of  the order parameter  and is independent of the
 sample size~\cite{mlbw99}.   The  collapse of these levels  on
 the ground-state is supposed  to be exponential with  the size of the
 lattice.

\item It is only when there is no long-range order, a
spin  gap and  when all spin  correlations are  short-ranged that  the
existence of  a ground-state degeneracy  is non-trivial. In that case,
the theorem  states  that the state(s)  degenerate   with the absolute
ground-state  in  the thermodynamic limit have  wave  vector $\pi$ with
respect to the ground-state.  More generally, in dimension larger than
1,  we will argue that the states which collapse to the ground-state
have wave vector ${\bf k}_{A_{i}} $ (wave vectors joining the
center  to     the   middle  of    the      sides of   the   Brillouin
zone)~\footnote{This result supposes that the ground-state wave vector
is zero.  It is  zero in all situations we  have been looking at. If a
system  had a non-zero momentum  ground-state,  it would be degenerate
and the theorem  true anyway !}.   These low lying  levels collapse to
the absolute ground-state exponentially with $L$.

\end{itemize}
\end{section}

\begin{section}{The LSMA variational state}
 
\begin{subsection}
{Finite size  spectra and Twisted Boundary Conditions.}

The spectrum of a given Hamiltonian on a  finite sample depends on the
boundary conditions: these can be free, periodic or twisted.  The LSMA
theorem is  valid  for periodic  boundary conditions.   For a complete
understanding of the nature of the LSMA variational state it is useful
to study how   a given exact spectrum  evolves  under a   twist of the
boundary conditions.

Let us consider a  finite sample in  $d$ dimensions, described  by $d$
vectors ${\bf    T}_j$. Generalized twisted  boundary   conditions are
defined by the   choice of the twist axis   (here the $z$ axis  in the
original ${\cal B}_0$ spin frame) and a set of $d$ angles $\phi_j$ as:

\begin{equation}
{\bf S}({\bf R}_i+{\bf T}_j)=e^{i\phi_jS^z({\bf R}_i)}
	{\bf S}({\bf R}_i)
	e^{-i\phi_jS_{z}({\bf R}_i)}
\label{twbc}
\end{equation}

From now on,   and for the sake   of simplicity, we will   develop the
algebra on  simple  1-dimensional   models (extrapolation to    larger
dimensions or more complicated $SU(2)$  invariant Hamiltonians is just
a problem  of notations).  As an example,  let us consider the nearest
neighbor Heisenberg Hamiltonian with periodic boundary conditions:

\begin{equation}
{\cal H}_0=\sum_{n=0}^{L-1} {\bf S}_n\cdot{\bf S}_{n+1 [L]}
\end{equation}
A twist $\phi$ in the boundary conditions
implies the calculation of the eigenstates of
\begin{eqnarray}
{\cal H}_{\phi}&=&{\bf S}_{L-1}\cdot
\left(e^{i\phi S^z_0}{\bf S}_0e^{-i\phi S^z_0}\right)+
\sum_{n=0}^{L-2} {\bf S}_n\cdot{\bf S}_{n+1} \nonumber \\
&=&{\cal H}_0
+\frac{1}{2} \left( ( e^{i\phi} -1) S_{L-1}^-S_0^++{\rm h.c.}\right)
\label{Hphi}
\end{eqnarray} 

Under an adiabatic twist of  the  boundary conditions the spectrum  of
${\cal  H}_\phi$  evolves periodically  with a period   $2\pi$, as the
boundary  conditions (Eq.~\ref{twbc}).  But  the eigenstates evolution
might  be more complicated: a  unique spin-$\frac{1}{2}$ wave function
acquires a phase  factor $-1$ under a $2\pi$  twist.  And  there is no
guaranty   that the     the  ground-state  of   ${\cal  H}_{\phi  =0}$
adiabatically transforms  into the ground-state  of ${\cal  H}_{\phi =
2\pi}$.  As we will show  below, this is  generally not the case and the
true period of the eigenstates is $4\pi$.

\end{subsection}

\begin{subsection}{Twisted Boundary  Conditions and Translational
Invariance }

To  follow adiabatically  an  eigenstate  while twisting  the boundary
conditions may  be difficult if  there are  level crossings during the
twist.  The  only way to  do it in an  unambiguous way is to  follow a
given one-dimensional representation  of the symmetry group during the
twist:  in such a representation   the levels are  non degenerate  and
never  cross.  The  ground-state   of  the Hamiltonian  with  periodic
boundary        conditions    indeed    belongs       to     such    a
representation~\footnote{Up to now, in any antiferromagnetic models we
have studied, the  ground-sate of an even  number of spins belongs  to
the trivial representation of the total group (the only exceptions are
associated to special pathological behavior of very small samples): it
is a  state with  total spin zero,   zero momentum,  invariant in  any
operation of the  point symmetry group.   This makes sense and appears
as a  powerful extension of Marshall  theorem  for bipartite lattices.
If  the ground-state      were    to belong   systematically   to    a
multidimensional  representation, as it might  be the  case for chiral
spin liquids, then the LSM theorem would again be trivially true.}.

Such a program is  not easy in  the framework of Eq.~\ref{Hphi}, which
breaks   the   translational   symmetry of   the     problem.  But the
translational  invariance can be restored  by a unitary transformation
rotating the spin frame   at each lattice  site.   Let us call  ${\cal
B}_{\phi}$, the new frame deduced from the original  ${\cal B}_0$ by a
spatially dependent twist described by the unitary operator:

\begin{equation}
 U(\phi)= \exp( i\frac{\phi}{L}   {\sum_{n=0}^{L-1} n  S^z_n})
\label{unit}
\end{equation}

In this new frame, the twisted Heisenberg Hamiltonian reads:
\begin{equation}
\tilde{{\cal H}}_\phi=U(\phi) {\cal H}_{\phi} U(\phi)^{-1};
\end{equation}
(in this equation and in the following, we put a tilde on each quantity
measured in the ${\cal B}_\phi$ frame).
This unitary transformation is chosen so that the boundary term(s) in
Eq.~\ref{Hphi} disappear(s):
\begin{equation}
\tilde{{\cal H}}_\phi=
{\sum_{n=0}^{L-1} \left[{S}_n^z{S}_{n+1 [L]}^z
+\frac{1}{2}\left(e^{i\phi/L}{S}_n^-{S}_{n+1 [L]}^++{\rm h.c.}\right)\right]}
\end{equation}
\begin{equation}
	\tilde{{\cal H}}_\phi=
	{\cal H}_{0}
	+\frac{1}{2}{\sum_{n=0}^{L-1}\left[(e^{i\phi/L}-1) {S}_n^-{S}_{n+1
[L]}^++{\rm h.c.}\right]}.
\label{h1h0}
\end{equation}
$\tilde{{\cal H}}_\phi$ is translation invariant ($ \left[\tilde{{\cal
H}}_\phi, {\cal  T}\right]=0$,  where ${\cal T}$   is the operator for
one-step translations)   and its spectrum  is  indeed identical to the
spectrum   of ${\cal   H}_\phi$.    We   can now define    irreducible
representations of the  translation    group labelled by   their  wave
vectors in the ${\cal B}_\phi$  frame and adiabatically follow a given
eigenstate of  the momentum in the  successive ${\cal  B}_\phi$ frames
while  increasing  $\phi$  from  $0$ to $     4 \pi$ (see  example  in
Fig.~\ref{twistedspectrum}).

For      a given   twist     $\phi$,    the zero-momentum   eigenstate
$\left|\tilde{\psi}_\phi ^{k=0}\right>$   of  $\tilde{\cal H}_\phi$ in
the  ${\cal B}_\phi$  frame has for   expression in  the  ${\cal B}_0$
frame:
\begin{equation}
\left|\psi_{\phi}^0\right>=U^{-1}(\phi)\left|\tilde{\psi}_\phi^{k=0}\right>.
\label{wavef}
\end{equation}
  For an arbitrary twist, $\left|\psi_{\phi}^0\right>$
 does not describe a spatially
homogeneous state in the ${\cal B}_0$ frame.

 We will now
show that for a $2 \pi$ twist, the trivial representation of
the translations in the ${\cal B}_{2 \pi}$ frame ($
\left|\tilde{\psi}_{2\pi}^{k=0}\right>$),  has 
momentum $   { \bf  k}_{A_i}$ in  the   ${\cal B}_0$ frame.  Following
Affleck \cite{a88}, this is easily done by noting that for odd-integer
spins $U(2\pi)$  anti-commutes with ${\cal T}$,  as soon as the number
of   rows     in     the    transverse     direction  is    an     odd
integer~\cite{lsm61,a88}.       This           proves             that
$\left|\psi_{2\pi}^0\right> $ defined by Eq. \ref{wavef} takes a phase
factor $-1$ in one-step translation along the twist direction and thus
has a momentum $ { \bf k}_{A_i}$ in the ${\cal B}_0$ frame.

An   example of   such     an adiabatic continuation   is    shown  in
Fig.~\ref{twistedspectrum}.  The spectrum  of the  multi-spin exchange
Hamiltonian  on  a small losange  ($  4 \times 5$)  is  displayed as a
function of  the twist angle.  The boundary  conditions are twisted in
the  direction of length  $L=4$ (the number  of rows is  odd).  We can
note the above-mentioned properties:
\begin{itemize}
\item The spectrum is periodic in $\phi$ of period $2\pi$.
\item The eigenstates of $\tilde{\cal H}_\phi$  and ${\cal T}$
evolve with a period
$4\pi$.
\item For a $2\pi$ twist, the zero momentum eigenstate of $\tilde{\cal
H}_{2\pi}$
in the ${\cal B}_{2\pi}$ frame has a momentum $ { \bf
k}_{A_i}$ in the ${\cal B}_0$ frame (compare the spectra and
labels for twists $0$ and $2\pi$)
\end{itemize}

\end{subsection}

\begin{subsection}{The LSMA variational state revisited}
The  proof of LSMA theorem  relies on the construction  of a low lying
excited state for the problem  with periodic boundary conditions.  Let
us call $\left|\psi_0\right>$ the exact  ground-state of this problem.
The LSMA   excited state is  obtained   by the  action of  the unitary
operator $U(2\pi)$ (Eq.~\ref{unit}) on $\left|\psi_0\right>$:
\begin{equation}
\left|\theta^{LSMA}_{2\pi}\right>=U(2\pi)\left|\psi_0\right>
\label{LSMA}
\end{equation}

As already mentioned, in specific geometries,
 $U(2 \pi)$ anti-commutes with ${\cal T}$
and $\left|  \theta ^{LSMA} _{2\pi}\right >$ has momentum $\pi$ with
respect to $\left| \phi _0\right>$. Contrary to the states described
above, $ \left |  \theta ^{LSMA} _{2\pi}\right >$ 
is not an eigenstate of
${\cal H}_0$. Its variational energy can be computed with
elementary algebra as:
\begin{equation}
\left< \theta ^{LSMA} _{2\pi}\right|
{\cal H}_0\left| \theta ^{LSMA} _{2\pi}\right>
=
\left<\psi_0\right|\tilde{{\cal H}}_{\phi=2\pi}\left|\psi_0\right>
.
\label{scalarproduct}
\end{equation} 

Using Eq.~\ref{h1h0} (generalized to d dimensions) we see that $\left|
\theta   ^{LSMA} _{2\pi} \right>$ variational  energy  is equal to the
exact    ground-state      energy       $\left<\psi_0\right|     {\cal
H}_0\left|\psi_0\right>$, plus a correction  of  the order of   ${\cal
O}(N\times(e^{2i\pi/L}-1))$, that  is ${\cal O}(N/L^2)$.   This gives,
in one dimension, an energy ${\cal  O}(1/L)$ and achieves the proof of
the LSM theorem.   The number of correction terms   does not allow  to
extend the proof to  higher  space dimensions~\footnote{ Only  systems
with aspect  ratios going to zero  as  the size  goes to infinity have
degenerate     states   reminding     of      the      one-dimensional
problem~\cite{a88}.}.

A   simple counter-example   showing   that $\left|    \theta  ^{LSMA}
_{2\pi}\right >$  is generally not a low  lying excitation is given by
the  spectra of the Heisenberg  hamiltonian on the triangular lattice:
in this situation  the exact  lowest  excitation energy (and thus  the
LSMA variational energy) in  the $ {  \bf  k}_{A_i}$ sector is  indeed
${\cal O}(1)$: it is associated to  the corresponding magnon which has
a    large  energy     in        this  three-sublattice        ordered
structure~\cite{bllp94}.

The  point  of view   developed in  the  previous subsection  provides
another way to look at the right-hand side of Eq.~\ref{scalarproduct}.
$\left|\psi_0\right>$ might be seen  as  a variational guess to  solve
the twisted  boundary conditions problem  using the $ {\cal B}_{2\pi}$
frame, and then  equivalently $\left| \theta ^{LSMA}  _{2\pi} \right>$
as the solution  of the  same physical  problem in the  ${\cal B}_{0}$
frame ($ {\cal H}_{\phi=2\pi} = {\cal  H}_0$).  This new point of
view enlights the  weakness of the  LSMA wave function (in $d>1$)  and
how we might try to overcome it.

Let us for the moment assume that $\phi$  is equal to $2\pi-\epsilon$.
In the $ {\cal B}_{0}$ frame, the perturbation to the periodic problem
only  appears  at a boundary of   dimension (d-1): but the variational
solution $\left| \theta ^{LSMA} _{2\pi - \epsilon} \right>$ smears out
the  spin  response on the entire  system.  This  would  be a sensible
solution  in  the case of  long  range order,  where  the system shows
stiffness   and sensitivity   to the  boundary   conditions.  On   the
contrary,  in  the present case, where   spin-spin correlations have a
finite range $\xi$,  it seems reasonable  to expect  that the boundary
perturbation  does not propagate at a  distance much larger than $\xi$
from the boundary.   The LSMA solution is  thus probably very far from
optimal.

We might thus expect to find in the $ {\cal  B}_{0}$ frame, a solution
with energy lower than  $\left|  \theta ^{LSMA} _{2\pi} \right>$ 
 by perturbing $\left|
\psi _{0}\right>$   only $locally$ at  the boundary.   For such a wave
function, Equation~(\ref{Hphi})    (and  its generalization    to  $d$
dimensions) implies a distance in energy  from the ground-state of the
order of  $ \epsilon^2 \,  \xi \,  L^{d-1}$.  We  could then speculate
that this  difference   in  energy might  be  tuned  to  zero   by  an
appropriate choice of the small free parameter $\epsilon$.  But indeed
such  a reasoning involves a difficult  and may be pathological limit.
In the following   paragraph we thus   propose a new  variational wave
function   for  the  low  lying  ${\bf  k}_A$ excited  state:  in  the
translationally broken  picture ($ {\cal  B}_{0}$), this excited state
differs very little from $\left|
\psi _{0}\right>$ and only in the vicinity of the boundary defect.

As  an   illustration of the  present  analysis,  we  can  look at the
evolution   of  the  low lying  levels   of  the spectrum of   the MSE
hamiltonian    under a   twist   $\phi$  of  the   boundary conditions
(Fig.~\ref{twistedspectrum}). In spite  of the very  small size of the
sample ($4\times 5$),  one clearly  sees  that the  exact ground-state
energy does not increase  as $\phi ^2$  but  more or less  as $\phi^4$
(see Fig.~\ref{twistedspectrumphi4}).  For  small enough $\phi$ , this
state does not present stiffness to  twisted boundary conditions: this
is exactly what is expected from a Spin-Liquid.

For comparison one can compute the variational energy of the state which
interpolates between $\left|\psi _{0}\right>$ and
$\left|\theta^{LSMA}_{2\pi}\right>$. It is defined as:
\begin{equation}
\left|\theta_{\phi}\right>=U^{-1}(\phi)\left|\psi_0\right>.
\label{LSM2}
\end{equation}
Its variational energy can  be rewritten  as  a linear combination  of
2-body and 4-body  observables of $\left|\psi  _{0}\right>$ multiplied
by cosines of  $\phi/L$. It thus increases as  $\phi ^2$ and has a non
zero stiffness,  which explains why it is  a bad  approximation of the
exact state.

Remark: Contrary to the  the $N=36$ spectrum of Fig.~\ref{sp36} (where
$ \xi > L$), the sample in Fig.~\ref{twistedspectrum} does not display
all  the  features   expected   from a    Spin-Liquid  spectrum.   The
correlations at  a distance  4  are still not  completely negligeable:
this explains the rapid increase in the  ground-state energy of ${\cal
H}_\phi$ for $\phi
\ge 0.2$.    As a consequence,  the  ordering of the   eigen-levels of
${\cal H}_0$ is different from  the thermodynamic limit: in particular
the ground-state in  the $ {\bf k}_A$ sector  is not the first excited
state of   the $4\times 5$ spectrum,  as  it is  in  the $6  \times 6$
example of Fig.~\ref{sp36}.  The evolution under a twist of the $N=36$
spectrum, would have been  much more pedagogic: it  is for the  moment
too expensive in computer time.

At that point it is interesting to discuss Oshikawa's approach of this
question~\cite{o99}.  In place  of  twisting the boundary  conditions,
Oshikawa suppose  that the quasi   particles  of the problem  have   a
fictitious charge. He inserts a magnetic flux in the hole of the torus
defined by  the  boundary conditions (see  Fig.~\ref{torus})  and then
increases  the flux  from   $0$  to  $2 \pi$.   This  is   a procedure
absolutely similar to our twist of the boundary conditions in the spin
problem,  and the  adiabatic  insertion of a  $2  \pi$ flux brings the
system   from     the     $\left|\psi_0\right>$  state        to   the
$\left|\tilde{\psi}_{k=0}\right>$ of  $\tilde{\cal H}_{2 \pi}$ defined
in subsection 3.2. At that point he does  the implicit hypothesis that
{\it the gap of the system does not close in  the operation}.  He thus
automatically arrives to   the  conclusion that  the  ground-state  is
degenerate.

Our approach gives  a complementary insight in  the physics of such  a
system: a Spin-Liquid, with its absence  of stiffness is not sensitive
to  a twist of the  boundary conditions  (in the thermodynamic limit).
Thus the  ground-state energy of  ${\cal H}_{\phi}$ in the translation
invariant sector does  not depend on the  twist $\phi$, which  implies
that the   ground-state of $   {\cal H}_0$  is  degenerate (with  wave
vectors $0$ and $  {\bf k}_{A_i}$ in the ${\cal  B}_0$ frame).  In the
following paragraph, using  a specific mathematical  definition of the
Spin-Liquid state, we  will exhibit a variational wave-function giving
a strong support to the LSMA conjecture.
\end{subsection}
\end{section}

\begin{section}{The short range RVB picture of the first excited states}

In this part we use the  main ideas of Sutherland~\cite{s88}, Read and
Chakraborty~\cite{rc89} to build an explicit variational wave-function
orthogonal   to the ground-state    and    collapsing to  it   in  the
thermodynamic  limit.  We  then show  that these  first excited states
have momentum ${\bf k}_{A_{i}}$ with respect to the ground-state.

Sutherland first showed that   the zero-temperature observables of   a
nearest neighbor resonating valence bond wave-function can be computed
thanks to the classical properties of a gas  of loops.  In the quantum
problem, the loops   appear  formally when   scalar  product of  wave-
functions  (or matrix elements   of spin permutations) are  written in
terms of dimer  coverings.  The Sutherland nearest neighbor resonating
valence-bond (NNRVB) wave-function  description  can be mapped  to the
high temperature  disordered phase  of  the classical  loop model: its
correlation length  is  finite   and  the weight   of long   loops  is
exponentially decreasing with their length.

Our own  reasoning rests on the  following {\bf basic assumption (A)}:
The ground-state $\left| \psi _{0}\right>$ is a  R.V.B. state, and the
long loops weight in the norm $\left<\psi_0\right|\left.\psi_0\right>$
decreases  exponentially with the  loop length.  This last requirement
implies {\it the  exponential decrease of all multi-point correlations
with distance}, the reverse  proposal  might equally  be true but  its
proof is less  obvious~\footnote{Miscellaneous remarks: i) a dimerized
state does  not   fulfill  property  (A) because it    has dimer-dimer
long-range order  (4-point correlation function).  ii) the Spin-Liquid
observed  on the kagom\'e  lattice might not obey  assumption A but it
has most probably gapless singlet excitations.  }.

The steps of the demonstration are as follows:
\begin{itemize}

\item Choice of a dimer basis.
Dimer decompositions are a bit uneasy  because dimer coverings are not
orthogonal    and the   entire     family  of   dimer  coverings    is
over-complete. So  we  must  suppose that  in   a first  time  we have
extracted a non orthogonal basis of independent dimer coverings called
generically     $\left|C\right>$.   This  basis  should  respect   the
translational  invariance  of   the   problem: which  means   that  if
$\left|C\right>$ is  a  basis vector,  so is  ${\cal T}\left|C\right>$
(where ${\cal T}$ is any unit step translation)

\item Decomposition of the translation invariant ground-state.
 We have understood previously that our problem is a boundary problem.
 Let us  draw a cut $\Delta$  encircling the torus created by periodic
 boundary  conditions  (see Fig.~\ref{torus}).   This hyper-surface of
 dimension $d-1$  cuts bonds  of  the lattice but  there  is  no sites
 sitting  on it.  The position  of  the cut  is arbitrary;  but we may
 decide  in order to follow closely  our previous discussion to put it
 between  spin $L-1$ and  spin 0 of  each row  of  the lattice. In the
 decomposition of $\left| \psi _{0}\right>$ on the dimer basis, let us
 sort  the   coverings in   two  sets  $\Delta_{+}$   and $\Delta_{-}$
 according to the parity $\Pi_{\Delta}$  of the number of bounds going
 across the cut.  This leads to a formal decomposition of $\left| \psi
 _{0}\right>$ in two parts:
\begin{equation}
\left| \psi_0\right> =\left|\psi_0^+\right> + \left|\psi_0^-\right>
\end{equation}
where the vectors  $\left|\psi_0^{\pm}\right>$  belong respectively to
the sets $\Delta_{\pm}$.

Let us now consider  a finite system with an  odd number of rows along
the  direction of   the   cut,  and a   one-step   translation  ${\cal
T}_{\Delta}$  that crosses the   cut.  If $\left|C\right>$ belongs  to
$\Delta_{+}$    ,  ${\cal     T}_{\Delta}\left|C\right>$ belongs    to
$\Delta_{-}$  and   reversely  (Property  (B)).     The  translational
invariance of $\left| \psi _{0}\right>$ thus implies that both $\left|
\psi _{0}^+\right>$ and $\left| \psi _{0}^-\right>$ are simultaneously
non zero.  With the assumption $(A)$, it is  easy to show that the two
components $\left|\psi_0^{\pm}\right> $ are orthogonal.
\item Excited state.
One can thus build the state
\begin{equation}
\left|\psi_{1,\Delta}\right> =\left|\psi_{0}^+\right> - \; \left|\psi_{0}^-\right>.
\end{equation}
Using  the fact that  $\left|  \psi  _{0}\right>$  has momentum  zero,
Property (B)  implies  that $\left|  \psi  _{1, \Delta}\right>$ has  a
momentum $k_{A}$ in  the direction of  ${\cal T}_{\Delta}$.   It seems
reasonable to think that   this property demonstrated  to be  true for
samples with an odd number of rows is valid in the thermodynamic limit
for any  samples   (this is  clearly  indicated by  small  size  exact
diagonalizations, see Fig.~\ref{sp36}  for  the $6 \times 6$  sample).
Remark: $\left|\psi_{1,\Delta}\right>$ is  obtained from $\left|  \psi
_{0}\right>$ by changing the sign of the dimers crossing the boundary:
in the case of a short range RVB state it is exactly the kind of local
perturbation we were searching for in the previous paragraph.
\item
The  demonstration  is   achieved   by  proving that    $\left|   \psi
_{1,\Delta}\right>$ and $\left| \psi  _{0}\right>$ have the same short
range   correlations and thus  the same   energy in the  thermodynamic
limit.    More precisely   only   the   phases   of the  {\em    long}
loops~\footnote{Loops with non-zero winding  number around the torus.}
in    the   expression   of   the    energy  $\left<\psi_0\right|{\cal
H}\left|\psi_0\right>$ change sign and thus  the variational energy of
$\left|\psi_{1,\Delta}\right>$ only  differs    from  the  energy   of
$\left|\psi_0\right>$    by  terms  of      the   order of  $    {\cal
O}(exp(-L/\xi))$  where   $\xi$ is  the  characteristic  length of the
loops.  Such a construction can be done  for any main direction of the
lattice, which  proves that the degeneracy of  the ground-state in the
thermodynamic limit is $2^d$  (a result already  obtained by  Read and
Chakraborty, without reference    to the wave   vectors  of the  quasi
degenerate ground-states)

\end{itemize}

This completes our assertion that the LSMA  conjecture is indeed valid
in a very  large number of  situations  whatever the dimension of  the
lattice.

\section{Miscellaneous remarks}
\begin{itemize}
\item
We thus arrive at the  conclusion that  in the  absence of long  range
order   there is, strictly   speaking,  symmetry breaking  of one step
translations.    Long range order  in  any  of these antiferromagnetic
systems  implies  symmetry breaking, BUT   the  reverse is  false, the
symmetry breaking described here does not imply  long range order in a
{\it  local} order parameter~\cite{mlbw99}   .   This property is   of
topological origin: the  observation  of this symmetry  breaking would
need sensitivity  to a  global  observable or  to  boundary conditions
(edge states).
\item
It should be noticed that the results on  the $6 \times 6$ sample seem
to indicate that the conserved symmetry of the ground-state is in fact
${\cal T}\Sigma$, where  $\Sigma$ is  the  reflexion  through a  plane
containing ${\bf k}_{A}$.
\end{itemize}

{\bf Acknowledgments:}  We acknowledge very  fruitful discussions with
S. Sachdev and C. Henley.  One of us (C.L.)  thanks the hospitality of
I.T.P.  and the organisors of the  Quantum Magnetism program.  Special
thanks are due to I.  Affleck and D.  Mattis whose vivid interest have
prompted  these remarks.  This research was  supported  in part by the
National Science  Foundation under Grant  No.  PHY94-07194, and by the
CNRS     and  the Institut   de   D\'eveloppement   des Recherches  en
Informatique Scientifique under contracts 994091-990076.
\end{section}
\begin{figure}
	\begin{center}
	\resizebox{11cm}{!}{
	\includegraphics{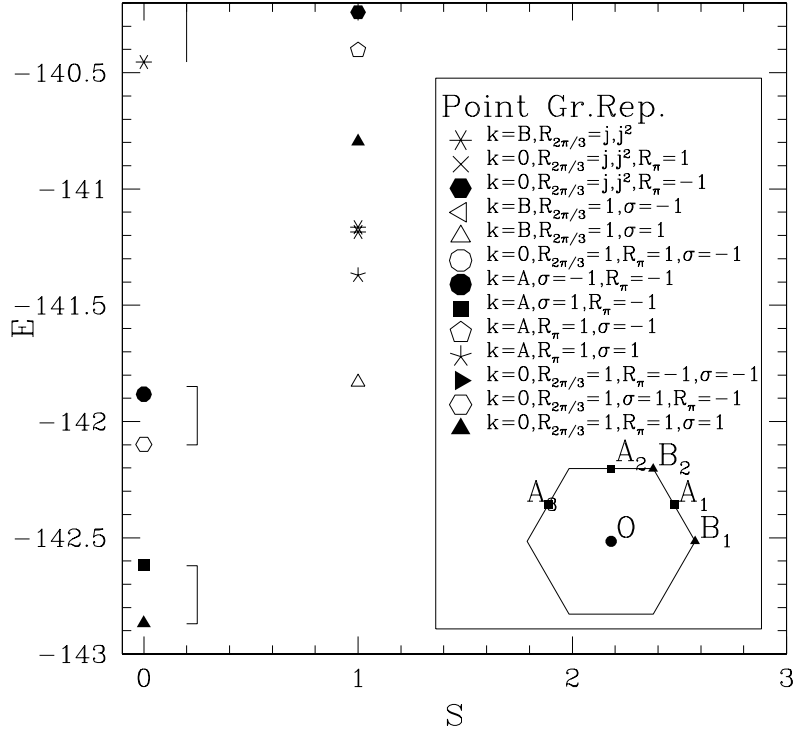}}
	\end{center}

	\caption[99] { First eigenstates of the multiple-spin exchange
	model on  a   $6 \times 6$  sites  sample (ref.~\cite{mlbw99},
	parameters  $J_2=-2$    and  $J_4=1$;  the  system is    in  a
	Spin-Liquid phase).  Eigenstates   with total spin   $S=0$ and
	$S=1$   are   displayed.  The  symbols   represent the spatial
	quantum numbers ( $\mathcal{R}_\theta$, and $\sigma=1$ are the
	phase  factors taken  by  the many-body   wave  function in  a
	rotation of the lattice of an angle $\theta$ or in a reflexion
	symmetry.).   The ground-state  ($E=-142.867$), belongs to the
	trivial representation of the space  group: ${\bf k}={\bf 0}$,
	$\mathcal{R}_{\frac    {2  \pi}{3}}=1$, $\mathcal{R}_{\pi}=1$,
	$\sigma=1$. The first excited  states have  wave-vectors  ${\bf
	k}_{A_i}$ (three-fold  degeneracy). The finite  size scaling
	strongly indicates that these states   collapse to the  absolute
	ground-state in the thermodynamic limit.  The third and fourth
	eigen-levels in the $S=0$ spin sector do probably not collapse
	to  the   absolute ground-state~\cite{mlbw99}.    They may  be
	degenerate in the thermodynamic  limit (4-fold degeneracy) and
	describe an   $S=0$ bound-state just  below   the continuum of
	triplet excitations.}\label{sp36}
\end{figure}
\begin{figure}
	\begin{center}
	\resizebox{10.5cm}{!}{
		\includegraphics{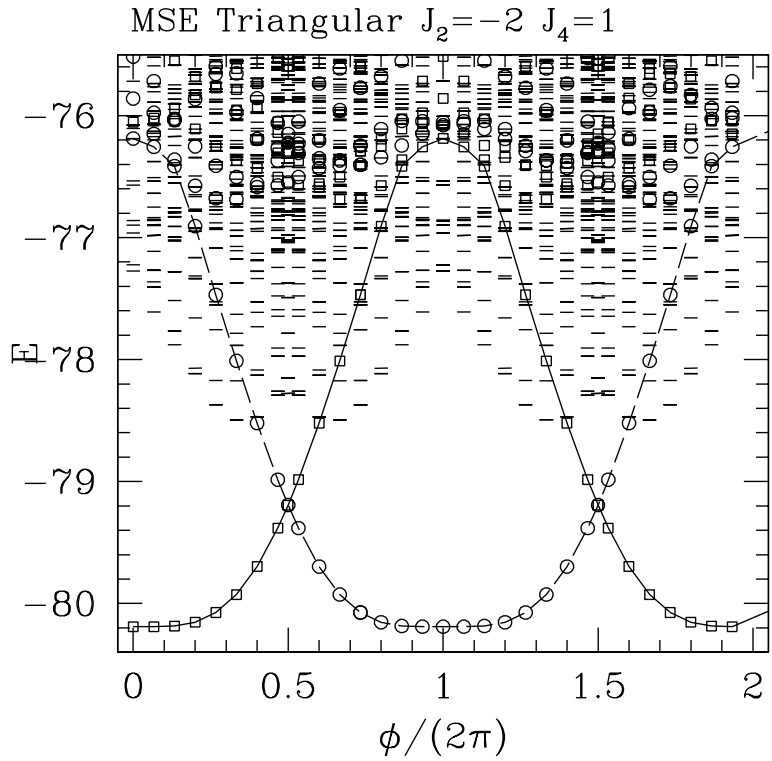}}
	\end{center}

	\caption{ Low-lying spectrum of the $4 \times 5$ sample of the
	multiple-spin    exchange  model  (same    parameters  as   in
	Fig.~\ref{sp36}), as a function  of the twist $\phi$.  Squares
	are ${\bf k}={\bf 0}$ states in  the ${\cal B}_\phi$ frame and
	circles   stand for states    with momentum  ${\bf  k}_{ A_1}$
	(momentum    $\pi$      in    the      even       direction).}
	\label{twistedspectrum}
\end{figure}

\begin{figure}
	\begin{center}
	\resizebox{8.5cm}{!}{
		\includegraphics{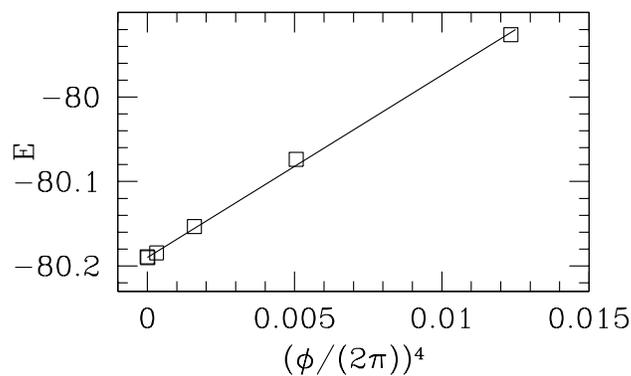}
		}
	\end{center}

	\caption{ Ground-state energy as a function of $\phi^4$ in the
	vicinity    of        $\phi=0$ (same    parameters  as in
	Fig.~\ref{twistedspectrum}).   The  energy  is  not  quadratic
	 but  rather  proportional  to
	$\phi^4$ (vanishingly  small   stiffness). 
     The  line   is   a guide    to   the eye.}\label{twistedspectrumphi4}
\end{figure}

\begin{figure}
	\begin{center}
	\resizebox{6cm}{!}{\includegraphics{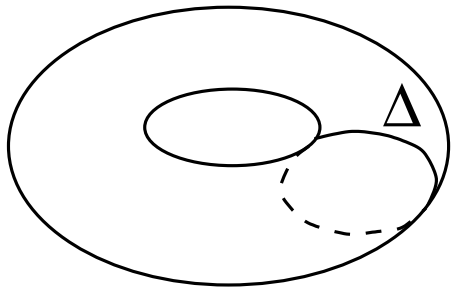}}
	\end{center} \caption{2-torus with one cut $\Delta$.} \label{torus} 
\end{figure}


\end{document}